\documentclass[11pt,a4paper]{article}
\pdfoutput=1
\usepackage{amsmath,amssymb,mathrsfs}
\usepackage{graphicx}
\usepackage{cite}

\usepackage{hyperref}
\hypersetup{bookmarks=true, bookmarksnumbered=true, linkcolor=blue,
citecolor=magenta,
urlcolor=blue,
colorlinks=true
}

\begin{document}
\begin{titlepage}
\begin{center}
\leavevmode \\
\vspace{-2 cm}

revised version \hfill UT-15-33 \\
October 2015 \hfill DESY 15-163 \\
\hfill IPMU 15-0143

\noindent
\vskip 3 cm
{\LARGE \bf Single-Superfield Helical-Phase Inflation}

\vglue.3in
{\large
Sergei V. Ketov~${}^{a,b,c}$ and Takahiro Terada~${}^{d, e}$ 
}

\vglue.1in

{\em
${}^a$~Department of Physics, Tokyo Metropolitan University \\
Minami-ohsawa 1-1, Hachioji-shi, Tokyo 192-0397, Japan \\
${}^b$~Kavli Institute for the Physics and Mathematics of the Universe (IPMU)
\\The University of Tokyo, Chiba 277-8568, Japan \\
${}^c$~Institute of Physics and Technology, Tomsk Polytechnic University\\
30 Lenin Ave., Tomsk 634050, Russian Federation \\
${}^d$~Department of Physics, The University of Tokyo, Tokyo 113-0033, Japan\\
${}^e$~Deutsches Elektronen-Synchrotron (DESY), 22607 Hamburg, Germany
}

\vglue.1in
ketov@tmu.ac.jp, takahiro@hep-th.phys.s.u-tokyo.ac.jp

\end{center}

\vglue.4in

\begin{abstract}
Large-field inflation in supergravity requires the approximate global symmetry needed to protect flatness of the scalar potential. In helical-phase inflation, the U(1) symmetry of the K\"ahler potential is assumed, the phase part of the complex scalar of a chiral superfield plays the role of inflaton, and the radial part is strongly stabilized. The original model of helical phase inflation, proposed by Li, Li and Nanopoulos (LLN), employs an extra (stabilizer) superfield.  We propose a more economical new class of the helical phase inflationary models without  a stabilizer superfield.  As the specific examples,  the quadratic, the natural, and the Starobinsky-type inflationary models are studied  in our approach.
\end{abstract}
\end{titlepage}


\section{Introduction}

Inflation well explains the origin of primordial density fluctuations, as well as flatness and homogeneity of our Universe.  The general idea is so far quite successful, and inflationary models are confronted with precise observational data~\cite{Ade:2015tva, Ade:2015lrj}.
Since inflation is a high-energy phenomenon, it is important to study it in a more fundamental framework such as supergravity~\cite{Ref:SUGRA} which is well motivated by particle physics and string theory. In particular, should tensor perturbations be detected in a near future, it would imply large (trans-Planckian) excursions of the inflaton field~\cite{Lyth:1996im}. Then the Planck-suppressed corrections cannot be neglected. Even if supersymmetry is broken at a higher scale than that of inflation, supergravity corrections have substantial impact on the scalar potential.

As is well known, a generic scalar potential in supergravity tends to be very steep in the large-field region, because of the exponential factor of the K\"{a}hler potential.  Accordingly, it is hard to achieve flatness of the scalar potential along the whole inflationary trajectory in the case of large-field models. Therefore, some symmetries are usually imposed in the inflationary model building in supergravity. A good example is the axion-like shift symmetry in the non-SUSY model~\cite{natural} and in the  supergravity-based models~\cite{KYY, Kallosh:2010ug}.

The similar approach, assuming the global U(1) symmetry and the related monodromy structure of the superpotential, is known as the 
helical-phase inflation, because its inflaton is identified with the phase component of a complex scalar field rolling down a helicoid 
potential~\cite{Li:2014vpa, Li:2014unh, Li:2015mwa, Li:2015taa}. Like in the more conventional shift-symmetric approach, a stabilizer superfield is used in the all known helical-phase inflationary models. Actually, the inflationary models with the non-compact U(1) and those based on the shift symmetry (with or without a stabilizer superfield) are equivalent, being related by the inflaton superfield redefinition $\Phi_{\rm U(1)}=\exp(\Phi_{\rm shift})$. Still, it makes sense to study them separately because adding a simple symmetry-breaking or stabilizing term in one approach often leads to a complicated structure in the other approach.  

In the shift-symmetric approach, a stabilizer superfield is needed to ensure positivity of the potential. In our previous 
work~\cite{Ketov:2014qha, Ketov:2014hya} (see also Ref.~\cite{Terada:2015sna}), we proposed the alternative framework to achieve the same goal by stabilizing the scalar superpartner of the inflaton.  In our approach, a number of the physical degrees of freedom was reduced, while a quartic shift-symmetric term was added to the K\"{a}hler potential.

In this letter, we study the helical-phase inflation without a stabilizer superfield. In Sec.~\ref{sec:stabilize} the radial part is stabilized by employing a higher-order (polynomial) term in the K\"{a}hler potential,  similarly to Refs.~\cite{Ketov:2014qha, Ketov:2014hya}. The inflaton is identified with the Pseudo-Nambu-Goldstone boson of the approximate U(1) symmetry of the K\"{a}hler potential. A few particular models are studied in Sec.~\ref{sec:models}. We conclude in Sec.~\ref{sec:conclusion}. Throughout the paper, we use the natural (reduced) Planck units,  $c=\hbar = M_{\text{P}}/\sqrt{8\pi}=1$.

\section{Stabilization of the radial component of complex inflaton} \label{sec:stabilize}

Should the inflationary trajectory be along the phase direction in the helical phase inflation, the radial direction has to be constant during inflation. For example, in the LLN model it is achieved by taking the superpotential proportional to a negative power of the inflaton. Balancing the superpotential contribution diverging at the origin with the exponentially rising contribution due to the K\"{a}hler potential results in the stabilization of the radial part at a value of the order of the Planck scale.

In our case without a stabilizer superfield, the inflaton potential includes both the superpotential itself and its derivative, and the formulae become rather complicated. Therefore, instead of dealing with a numerical minimization of the radial part, we employ the strong stabilization mechanism by using a higher order term in the K\"{a}hler potential,
\begin{align}
K=\left( \bar{\Phi}\Phi-\Phi_0^2 \right) - \frac{\zeta}{4}\left( \bar{\Phi}\Phi -\Phi_0^2 \right)^4. \label{Kahler}
\end{align}
The first term is the usual (minimal) K\"{a}hler potential.
The constant term is added so that the expectation value of the K\"{a}hler potential approximately vanishes.
The second term is introduced for the purpose of stabilization.
Thanks to that term, the radial part is stabilized at 
\begin{align}
|\Phi| \simeq  \Phi_0 - \frac{\Phi_0 \left( \Phi_0^2-2 \right)}{12\zeta \Phi_0^6 + 2\Phi_0^4 -\Phi_0^2-2}~~. \label{radVEV}
\end{align}
This expression is obtained by expanding the potential up to the second order in $|\Phi|-\Phi_0$ and minimizing it.
We assume the potential can be approximated as $V= (K^{\Phi}K_{\Phi}-3) |W|^2 = (\Phi_0^2 -3 ) |W|^2$, and neglect derivatives of the superpotential since they are proportional to the slow-roll parameters. 
 As expected, the stabilized value of the radial part approaches to $|\Phi|=\Phi_0$ as $\zeta$ goes to infinity. Moreover, $\zeta$ of order one or smaller is sufficient for the truncation to be consistent in the case of $\Phi_0>1$.
More general K\"{a}hler potentials with similar features may exist, but we find the above example to be simple and efficient.
Our stabilization mechanism is similar to those considered in the literature~\cite{Ellis:1984bs, Lee:2010hj, Kallosh:2013xya, Ketov:2014qha, Ketov:2014hya, Terada:2015sna}.

The stabilization parameter $\zeta$ in eq.~\eqref{Kahler} should be real and positive. Some comments about its magnitude are in order.
When $\zeta$ becomes large at a fixed $\Phi_0$, the K\"{a}hler metric (the coefficient at the kinetic term) may change its sign before reaching the symmetric phase, $\langle \Phi \rangle =0$.  It occurs when $\zeta > 4/\Phi_0^6$ for the K\"{a}hler potential in Eq.~\eqref{Kahler}. Then the above K\"{a}hler potential should be regarded as the effective description of the Higgsed phase around $\langle \Phi \rangle \sim \Phi_0$. It is enough for our purposes, since the radial part is stabilized throughout the process of inflation so we do not have to consider its dynamics.\footnote{\label{fn:AwayFromMinimum}
When using the terms $(\bar{\Phi}\Phi - \Phi_0^2 )^n$ with $n\neq 1, \, 4$, also allowed by the symmetry, in Eq.~\eqref{Kahler}, dynamics of the radial part cannot be predicted once its distance from $\Phi_0$ is more than that of the order one. It may cause the typical problem of initial conditions for inflation. We just assume here that the radial part is within the order-one distance from $\Phi_0$ at the onset of inflation.
}  Conversely, if $\zeta$ becomes small at a fixed $\Phi_0$, the stabilized position of the radial part shifts inwards, $|\Phi|< \Phi_0$, and eventually moves to the origin for $\zeta \to 0$.
Depending on the value of $\zeta$, 
it is caused either by classical inflaton dynamics, quantum fluctuations, or quantum tunneling. To avoid
such situations, we take the value of $\zeta$ to be of at least the same order as that of the critical value $4/\Phi_0^6$.

The stabilization strength can be measured by the mass of the radial part.  When $|\Phi|=\Phi_0$, the canonically normalized squared mass of the radial part is given by
\begin{align}
m_{\text{radial}}^2 \simeq  \frac{3(12 \zeta \Phi_0^6 +2 \Phi_0^4 - \Phi_0^2-2)}{\Phi_0^2 -3} H^2  \gtrsim 20 H^2\label{rad_mass}
\end{align}
under the same conditions used for deriving eq.~\eqref{radVEV}. In the last inequality we also assume that $\zeta \geq 0$ and $\Phi_0^2>3$. Therefore, it is not difficult to strongly stabilize the radial part, \textit{i.e.} $m_{\text{radial}}>H$. As long as the radial component is stabilized with its mass much larger than the Hubble scale, the considerations in the next Sections are independent upon the detailed mechanism of the stabilization.

After inflation, we cannot neglect the derivatives of the superpotential, so that Eqs.~\eqref{radVEV} and \eqref{rad_mass} are no longer valid. The model-independent expression for the shift of the radial value from $\Phi_0$ is very complicated, but it vanishes in the limit $\zeta \to \infty$. The radial component is not protected by any symmetry, contrary to the phase component.  We expect the mass of
the radial component to be of at least the same order as the mass of the phase component.  As will be clear in Section~\ref{sec:models}, SUSY is generically broken spontaneously.  There is a SUSY breaking mass contribution $6\sqrt{\zeta} \Phi_0^2 m_{3/2} $ near the vacuum, which becomes dominant in the large $\zeta$ limit. Thus, the radial part is kept fixed near $\Phi_0$ after inflation for a sufficiently large value of $\zeta$.

\section{Helical phase inflationary models in our approach} \label{sec:models}
Having stabilized the radial mode at $|\Phi| = \Phi_0$, let us consider typical inflationary models, without introducing a stabilizer superfield.
Let us parametrize the inflaton field as $\Phi=\Phi_0 e^{i \theta / \sqrt{2} \Phi_0}$.
The phase is scaled so that it is canonically normalized.
The superpotential breaks the U(1) symmetry in the K\"{a}hler potential, and generates the inflaton (scalar) potential.
We study chaotic inflation with the quadratic potential, the Starobinsky-like plateau potential, and a sinusoidal potential 
in this Section.

\subsection{Quadratic helical-phase inflation} \label{ssec:quadratic}
The logarithmic singularity in the superpotential is the heart of the helical phase inflation,
which is needed to realize a nontrivial spiral shape.
Let us take the simplest Ansatz 
\begin{align}
W=m \log \frac{\Phi}{f}~~,
\end{align}
where $m$ sets the scale of inflation, and $f\equiv f_0 e^{i \theta_0 / \sqrt{2} \Phi_0}$ (with $f_0$ and $\theta_0$ real) is the dimensional parameter controlling the cosmological constant.

\begin{figure}[tbhp]
  \begin{center}
    \includegraphics[clip,width=10.6cm]{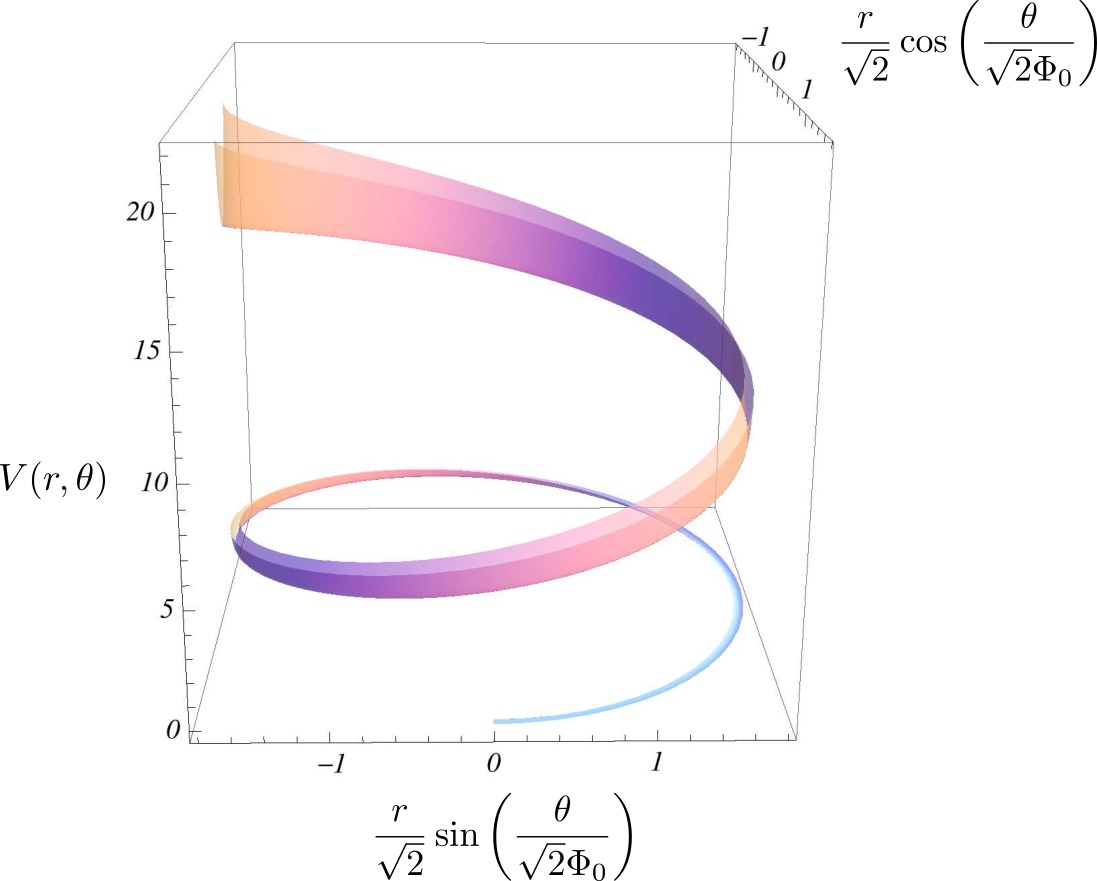}
    \caption{The quadratic potential for helical phase inflation.  The parameters are chosen as $\Phi_0=1.8$, $\zeta=0.11$, and $m=1$. The complex scalar is parametrized as $\Phi= \frac{r}{\sqrt{2}} e^{i\theta/ \sqrt{2}\Phi_0}$.}
    \label{fig:quadratic}
  \end{center}
\end{figure}

After stabilization, the inflaton potential becomes
\begin{align}
V = \frac{1}{2}m_{\text{inf}}^2 \left( \theta - \theta_0 \right)^2 + \Lambda,
\end{align}
with
\begin{align}
m_{\text{inf}}=& \frac{|m| \sqrt{\Phi_0^2-3}}{\Phi_0}~~, \\
\Lambda = & |m|^2 \left(  \frac{1}{\Phi_0^2} + 2 \log \frac{\Phi_0}{f_0} + \left( \Phi_0^2 -3 \right) \left| \log \frac{\Phi_0}{f_0} \right|^2 \right).
\end{align}
Thus, the positive quadratic scalar potential is obtained under the condition $\Phi_0 >\sqrt{3}$.
The cosmological constant can be eliminated by choosing
\begin{align}
f_0 = \Phi_0 e^{\frac{1}{\Phi_0 (\Phi_0 \pm \sqrt{3})}}.
\end{align}
The full potential is shown in Fig.~\ref{fig:quadratic} for a limited field range.
As is clear from the Figure, the radial part is strongly stabilized, while its mass increases with the potential.
This is also implied by Eq.~\eqref{rad_mass}.

In this model, SUSY is not restored after inflation.
The gravitino mass at the vacuum is
\begin{align}
m_{3/2}=\frac{|m|}{\Phi_0 ( \Phi_0 \pm \sqrt{3} ) }~~.
\end{align}
On the one hand, in the case of $(\Phi_0 -\sqrt{3} )\ll 1 $, the inflaton becomes much lighter than the gravitino, $m_{\text{inf}}\ll m_{3/2}$. On the other hand, in the large $\Phi_0$ limit, the inequality is reversed, $m_{3/2}\ll m_{\text{inf}}$. 

The inflationary predictions of the model are well-known, and they do not change in our embedding.
Nevertheless, a few comments are in order.
By using the observed value of the amplitude of scalar perturbations, $A_{\text{s}}=2.2\times 10^{-9}$, the inflaton mass is determined as $m_{\text{inf}}=1.8 \, (1.5)\times 10^{13}$ GeV at the $e$-foldings $N=50\, (60)$. The gravitino mass is roughly of the same order.
SUSY breaking is then supposed to be mediated to the MSSM sector by the anomaly mediation, and is not compatible with the traditional low-energy SUSY scenario. It is the typical consequence of removing a stabilizer field, see Ref.~\cite{Ketov:2014qha}. In the last case, SUSY breaking at the intermediate scale~\cite{Hall:2009nd, Hall:2013eko, Hall:2014vga, Ibanez:2012zg, Ibanez:2013gf, Hebecker:2012qp,  Hebecker:2013lha} can be motivated \textit{e.g.}, by noticing that it stabilizes the electro-weak vacuum.

Though the quadratic potential is already excluded by Planck observations, some modifications or coupling to other sectors may make the quadratic model consistent with the data (see \textit{e.g.,} Ref.~\cite{Buchmuller:2015oma}). Instead of studying such possibilities, we directly construct some viable inflationary models in the next subsections.

\subsection{Starobinsky-like helical-phase inflation} \label{ssec:plateau}
In the previous Subsection, the logarithm $\log \Phi = \log \Phi_0 + i\theta / (\sqrt{2}\Phi_0)$ in the superpotential leads to the quadratic potential.
A plateau-type potential consists of the exponential factors like $e^{-\theta}$, so let us consider the exponential of the logarithm, $e^{i \log \Phi} = \Phi^{i}$ which is equivalent to the imaginary power of the superfield. In other words, let us take the following superpotential:
\begin{align}
W=m \left( c + \Phi^i \right), \label{Wplateau}
\end{align}
where $m$ and $c$ are the constant parameters that determine the scale of inflation and the cosmological constant (see
below).

After stabilization, the inflaton potential becomes
\begin{align}
V=|m|^2 \left(  A+B e^{-\theta/\sqrt{2}\Phi_0} + C e^{-2 \theta/ \sqrt{2}\Phi_0} \right),
\end{align}
with the coefficients
\begin{align}
A=& |c|^2 \left( \Phi_0^2 -3 \right), \\
B=& 2 |c| \left[ \left( \Phi_0^2 -3 \right) \cos \left( \log \Phi_0 -\varphi \right) - \sin \left( \log \Phi_0 -\varphi \right) \right] , \\
C=& \Phi_0^2 -3 +\frac{1}{\Phi_0^2}~~,
\end{align}
where the phase $\varphi$ is defined by $c=|c| e^{i \varphi}$.

\begin{figure}[tbhp]
  \begin{center}
    \includegraphics[clip,width=10.6cm]{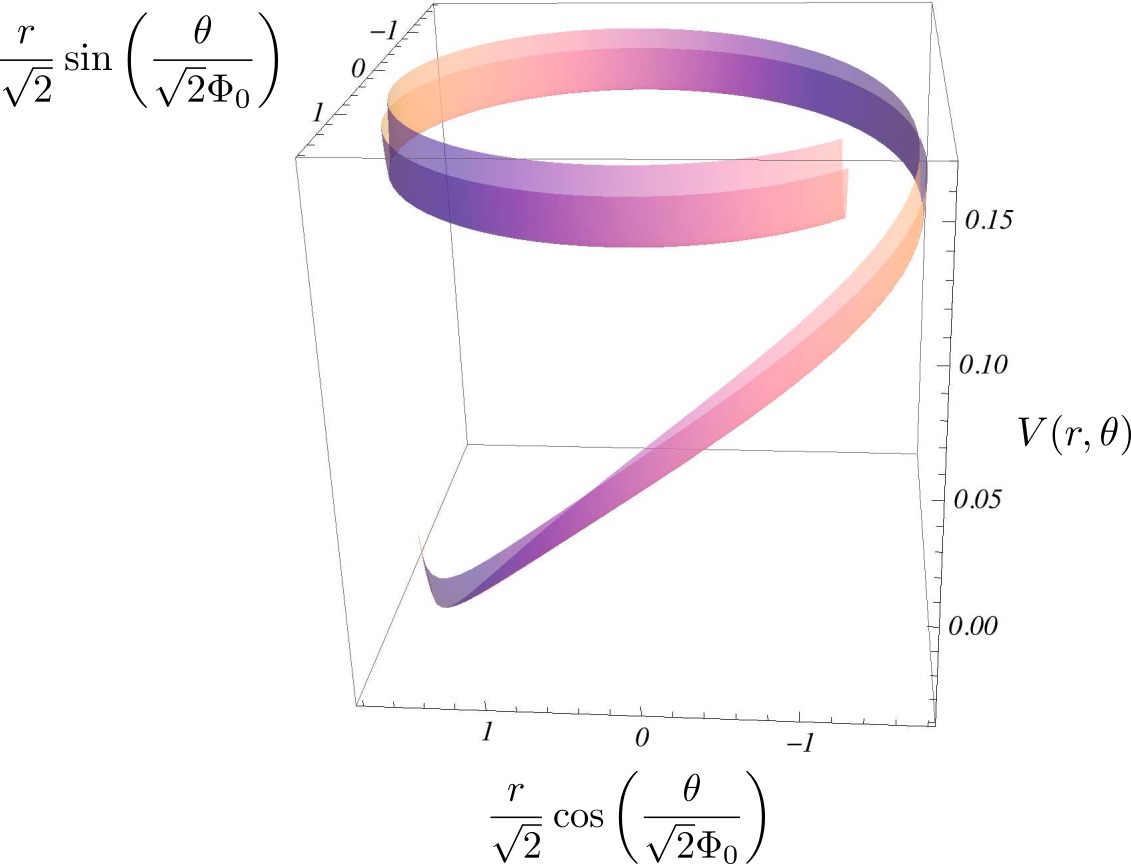}
    \caption{The Starobinsky-like potential for helical phase inflation.  The parameters are chosen as $\Phi_0=1.8$, $\zeta=0.11$, $\varphi=3.85$, and $m=1$. The complex scalar is parametrized as $\Phi= \frac{r}{\sqrt{2}} e^{i\theta/ \sqrt{2}\Phi_0}$.}
    \label{fig:Starobinsky}
  \end{center}
\end{figure}

For any $\Phi_0$ larger than $\sqrt{3}$, $A$ and $C$ are positive definite, and the sign of $B$ depends on $\varphi$.
There exists a solution of $\varphi$ such that $B$ is negative and, moreover, the cosmological constant vanishes.
Though it is fully straightforward, the solution itself is not very illuminating and, hence, is not shown here.
The potential is a generalization of the Starobinsky potential.  Such potentials are often called  ``Starobinsky-like'' in the literature. Our Starobinsky-like scalar potential is shown in Fig.~\ref{fig:Starobinsky}.

The masses of inflaton and gravitino are given by
\begin{align}
m_{\text{inf}}=&\frac{|m c | }{\Phi_0}\sqrt{\Phi_0^2 -3}~~, \\
m_{3/2}=& |mc| \left| e^{i\varphi}- e^{i\log\Phi_0}\sqrt{ \frac{\Phi_0^2 (\Phi_0^2-3)}{\Phi_0^2 (\Phi_0^2-3)+1} }  \right|.
\end{align}
The inflaton mass reads $m_{\text{inf}}=3.5\, (2.9)\times 10^{13}$ GeV at $N=50 \, (60)$.
In this model, the inflaton is always lighter than the gravitino.  In the limit $\Phi_0 \to \infty$,
the inflaton mass approaches half of the gravitino mass. 

The spectral index is the same as that of the Starobinsky model, but the tensor-to-scalar ratio is different:
\begin{equation}
1-n_{\text{s}}= \frac{2}{N} \qquad {\rm and}  \qquad
r= \frac{16 \Phi_0^2}{N^2}~~, 
\end{equation}
in the leading order of $N^{-1}$.
With $\Phi_0^2>3$, the tensor-to-scalar ratio is enhanced, when being compared to the Starobinsky model ($r=12/N^2$).
With an arbitrary imaginary power $\Phi^{bi}$ instead of $\Phi^i$ in Eq.~\eqref{Wplateau}, where $b$ is a real parameter, the tensor-to-scalar ratio is divided by $|b|$ as $r=16\Phi_0^2 / |b| N^2$.

\subsection{Natural helical-phase inflation} \label{ssec:natural}

The previous examples are based on the superpotentials having the singularity at the origin. However, it is not the necessary feature of our mechanism because of the super-Planckian value of the radial component. Let us take the superpotential
of the previous Subsection and replace its imaginary power by a real power as
\begin{align}
W=m\left( c+ \Phi \right).
\end{align}
This is simply a linear function without the monodromy structure.
In this case, a large value of $|\Phi|$ is required not only by the positivity of the stabilized potential but also by the observational status of natural inflation.

\begin{figure}[tbhp]
  \begin{center}
    \includegraphics[clip,width=10.6cm]{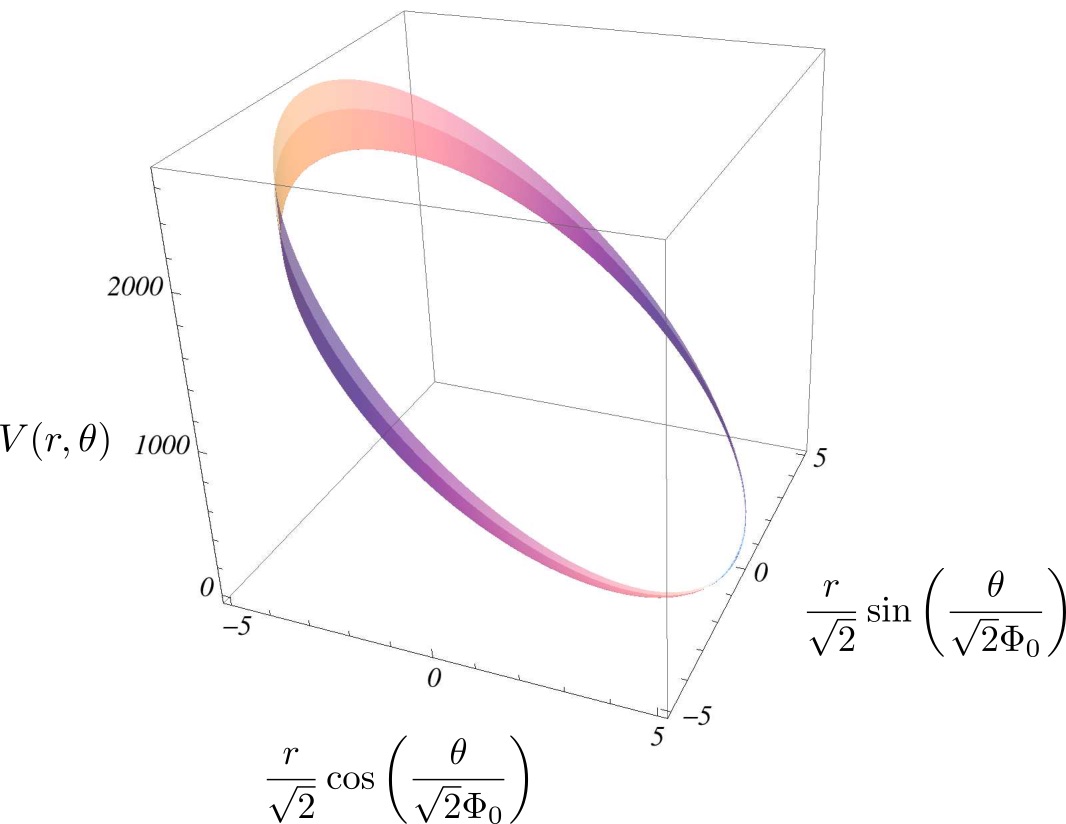}
    \caption{The sinusoidal potential for helical phase inflation.  The parameters are chosen as $\Phi_0=5$, $\zeta=0.03$, $m=1$, and $\varphi=\pi$, and the upper sign is taken in Eq.~\eqref{c_natural}. The complex scalar is parametrized as $\Phi= \frac{r}{\sqrt{2}} e^{i\theta/ \sqrt{2}\Phi_0}$.}
    \label{fig:natural}
  \end{center}
\end{figure}

After stabilization, the inflaton potential becomes
\begin{align}
V=|m|^2 \left[ D+ E \cos \left( \frac{\theta}{\sqrt{2}\Phi_0} - \varphi \right) \right]~,
\end{align}
with the coefficients
\begin{align}
D=&|c|^2 (\Phi_0^2-3) + \Phi_0^4 -\Phi_0^2 +1~~ ,\\
E=&2|c| \Phi_0 (\Phi_0^2-2)~~,
\end{align}
and $\varphi$ is again the argument of $c$, $c=|c|e^{i\varphi}$.
The cosmological constant vanishes when 
\begin{align}
|c| = \frac{\Phi_0 (\Phi_0^2-2) \pm \sqrt{3}}{\Phi_0^2-3}~~. \label{c_natural}
\end{align}
In this case, the potential is positive when $\Phi_0^2 >3$ (2) for the upper (lower) sign, and
the sinusoidal scalar potential of natural inflation is obtained.
The potential is shown in Fig.~\ref{fig:natural}.

The masses of inflaton and gravitino are given by
\begin{align}
m_{\text{inf}}=&|mc| \sqrt{\frac{\Phi_0^2-2}{\Phi_0}} ~~, \\
m_{3/2}=& \frac{|m|}{\Phi_0 \mp \sqrt{3} }~~.
\end{align}
Again, if the absolute value of the field is barely larger than the critical value $\sqrt{2}$ (this is for the lower sign), the inflaton is much lighter than the gravitino.  In the large VEV case, gravitino becomes much lighter than the inflaton.

The parameter of the natural inflation is tightly constrained by the CMB observations.  The decay constant (in our case $\sqrt{2}\Phi_0$) must be larger than $6.9$ at 95\% confidence level~\cite{Ade:2015lrj}, so that the lower bound on the absolute value is obtained as $\Phi_0 \gtrsim 4.9$.
When choosing $\Phi_0=5$, the inflaton mass is $m_{\text{inf}}=1.1\, (0.96) \times 10^{13}$ GeV at
 $N=50 \, (60)$.


\section{Conclusion} \label{sec:conclusion}

In this paper we studied helical phase inflation with a single chiral superfield in supergravity, \textit{i.e.} without
the stabilizer superfield used in the known versions of helical phase inflation in the literature.

In order to ensure positivity of the scalar potential and avoid computational complexity, we introduced 
a stabilization term to the K\"{a}hler potential that fixes the radial component of the inflaton complex scalar 
at a sufficiently large value. It results in technical simplification also. After the stabilization, a slow-roll inflation 
occurs in the direction of the phase component.

We exemplified our findings on the three simple models of the single-superfield helical-phase inflation.
It implies that there should be many more possibilities to obtain viable inflationary potentials in our approach.
One such noticeable generalization is a hybrid version of the models in Subsections~\ref{ssec:plateau} and \ref{ssec:natural}.  Let us take an arbitrary complex power of the inflaton superfield, $W=m(c+\Phi^{a+ib})$ with $a$ and $b$ real.
This model interpolates between the natural inflation and the Starobinsky-like inflation.
A similar model was studied in the presence of the stabilizer superfield in Ref.~\cite{Li:2015taa}.

Although the radial part is stabilized at an over-Planckian value, higher order terms in the K\"{a}hler potential are not necessarily problematic as long as we expand it around $\bar{\Phi}\Phi=\Phi_0^2$ as in eq.~\eqref{Kahler} (see also footnote~\ref{fn:AwayFromMinimum}).
However, possible shift symmetry breaking terms in the K\"{a}hler potential may affect the inflaton potential and inflationary observables~\cite{Price:2015xwa, Flauger:2009ab, Li:2015mwa}.
It is an interesting topic to be studied also in our setup elsewhere.

In the models in Subsections \ref{ssec:quadratic} and \ref{ssec:plateau}, singularities and the monodromy structure are introduced in the superpotential, as in the LLN approach.
This is beyond the usual field theory framework, and it is regarded as an effective description.
A possible UV completion of helical phase inflation in string theory was argued in Ref.~\cite{Li:2015taa}.

In summary, we proposed the new type of inflationary mechanism in supergravity, combining the ideas of helical phase inflation~\cite{Li:2014vpa, Li:2014unh, Li:2015mwa, Li:2015taa} and single-superfield inflation with the higher dimensional stabilization term in the K\"{a}hler potential~\cite{Ketov:2014qha, Ketov:2014hya}.  Our models are simple: the kinetic term is approximately canonical, the superpotential is very economical, and no stabilizer superfield (or extra d.o.f.) is present.

\section*{Acknowledgements} 
The authors are grateful to the referee for his/her careful reading of the manuscript and critical remarks.
SVK is supported by a Grant-in-Aid of the Japanese Society for Promotion of Science (JSPS) under No.~26400252, the World Premier International Research Center Initiative, MEXT, Japan, the TMU Special Research Fund and the Competitiveness Enhancement Program of the Tomsk Polytechnic University in Russia. SVK thanks the CERN Theory Group for kind hospitality extended to him during preparation of this paper.  TT acknowledges the JSPS for a Grant-in-Aid for JSPS Fellows and the Grant-in-Aid for Scientific Research No.~26$\cdot$10619.


\end{document}